\documentclass[12pt]{iopart}
\usepackage{iopams}  
\usepackage{graphicx,epsfig}
\usepackage{hyperref}
\newcommand{\beq}{\begin{equation}}
\newcommand{\eeq}{\end{equation}}
\newcommand{\ec}{\end{center}}
\newcommand{\bc}{\begin{center}}
\newcommand{\eea}{\end{eqnarray}}
\newcommand{\bea}{\begin{eqnarray}}
\newcommand{\eeas}{\end{eqnarray*}}
\newcommand{\beas}{\begin{eqnarray*}}
\newcommand{\bd}{\begin{description}}
\newcommand{\ed}{\end{description}}
\newcommand{\vx}{\mathbf{x}}
\newcommand{\vxp}{\mathbf{x}'}
\newcommand{\ei}{\end{itemize}}
\newcommand{\vk}{\mathbf{k}}
\newcommand{\vko}{\mathbf{k}_1}
\newcommand{\vkt}{\mathbf{k}_2}

\newcommand{\vkth}{\mathbf{k}_3}

\newcommand{\vp}{\mathbf{p}}
\newcommand{\vq}{\mathbf{q}}

\newcommand{\mysubsection}[2][]

\begin{document}
\title{Steady-state skewness and kurtosis from renormalized cumulants in $(2+1)$-dimensional stochastic surface growth}
\author{Tapas Singha \&  Malay K. Nandy}
\address{Department of Physics, Indian Institute of Technology Guwahati, Guwahati 781039, India.}
\ead{s.tapas@iitg.ernet.in \& mknandy@iitg.ernet.in}
\vspace{10pt}
\date{August 26, 2016}

\begin{abstract}
 The phenomenon of stochastic growth of a surface on a two-dimensional substrate occurs in Nature in a variety of circumstances and its statistical characterization requires the study of higher order cumulants. Here, we consider the statistical cumulants of height fluctuations governed by  the $(2+1)$-dimensional KPZ equation for flat geometry. We follow a diagrammatic scheme to derive the expressions for renormalized cumulants up to fourth order in the stationary state. Assuming a value for the roughness exponent from reliable numerical predictions, we calculate the second, third and fourth cumulants, yielding skewness $S=0.2879$ and kurtosis $Q=0.1995$.  These values agree well with the available numerical estimations.  
\end{abstract}
\hspace{4pc}

\section{Introduction}
The scale invariant growth of a two dimensional surface is a subject of great significance in nonequilibrium statistical mechanics. 
This is due to its wide range of applicability in addition to its theoretical complexity \cite{book_stanley,krug97,Halpin95,Meakin93,Family_Physica}. 
It has almost been three decades that Kardar, Parisi, and Zhang \cite{KPZ86} proposed a generic equation for surface growth, namely, 
\beq
\frac{\partial h} {\partial t}=\nu_0
\nabla^{2} h+\frac{\lambda_0}{2} (\nabla h)^{2}+ \eta,
\label{eq-kpz}
\eeq
known as the KPZ equation, where $h(\vx,t)$ is the fluctuating height field and $\nu_0$ is the surface tension. 
The surface grows due to aggregation of particles, modeled by the stochastic noise term $\eta(\vx,t)$, which is considered to be 
Gaussian of zero average with the correlation  
\beq
\langle\eta(\vx,t)\eta(\vxp,t')\rangle=2D_{0}\delta^d(\vx-\vxp)\delta(t-t'),
\label{eq-noise-2D}
\eeq
where $D_0$ is referred to as the deposition noise strength and $d$ is the dimension of the substrate. 

The  KPZ equation in one dimension plays an important role in the application domain. 
The $(1+1)$-dimensional KPZ equation has a semantic relation to a variety of systems. For example, directed polymers in random media 
(DPRM) \cite{kpz87,Fisher_PRB_43_10728}, vorticity free fluid velocity described by the Burgers equation \cite{forster_pra_16_732_77}, 
the stochastic heat equation (SHE) \cite{Healy13_PRE.88.042118}, and even sequence alignments in proteins and genes 
\cite{Hwa_Lassig_PRL_76_2591,Hwa_gene_allignment_nature_399}, growth phenomena in bacteria colonies \cite{Vicsek_90,Huergo_10}, turbulent 
liquid crystals (TLC) \cite{K_A_Takeuchi,kazumasa},
slow combustion of a sheet of paper \cite{Maunuksela97-PRL.79.1515,Myllys_PRE.64.036101,miettinena_epjb_46_55_05}, etc., exhibit 
the same scaling exponents as the $(1+1)$-dimensional KPZ equation.

The KPZ equation has been analyzed through renormalization group \cite{KPZ89,frey_pre_50_1024_94},
mode coupling calculation \cite{Beijeren_PRL_54_2026,huse85,Frey96_PRE.53.4424}, and numerical simulations 
\cite{jmkim_prl_62_2289_89,Forrest-Tang90_PRL.64.1405,Amar-Family-PRA.41.3399}. Moreover, the scaling functions \cite{Hwa_PRA_44_R7873}, 
as well as scaling exponents and the probability distribution function have been studied 
through finite temperature DPRM in $(1+1)$ dimensions \cite{Healy91-PRA.44.R3415} and zero temperature 
DPRM in $(1+1)$ and $(2+1)$ dimensions \cite{JMKIM}.

A considerable amount of understanding of the probability distribution function and its dependence on the initial conditions 
(namely flat, curved, and stationary) has been achieved through the study of various analytical \cite{calabrese_prl_106_250603_11,Imamura_prl_108_2012},  
numerical \cite{Michael_Herbert_00} and experimental \cite{K_A_Takeuchi,kazumasa,Takeuchi12} methods for different systems that are governed by the 
$(1+1)$-dimensional KPZ type dynamics. The evolution of a growing surface from a flat initial condition to the stationary state 
has been studied \cite{Takeuchi_prl_110_2013} both numerically (PNG) and experimentally (TLC). The corresponding crossover function is established 
as universal \cite{Healy_Lin_PRE.89.010103} by considering DPRM, stochastic heat equation (SHE) and growth models which share the universality class 
of the $(1+1)$-dimensional KPZ dynamics.

There exists no exact solution for the $(2+1)$-dimensional KPZ equation, which represents a wide variety of surface growth phenomena in real life. 
Recently the $(2+1)$-dimensional KPZ has been realized as an important problem where the higher dimensional analogs of TW GOE, TW GUE and 
Baik-Rains distribution have been investigated \cite{Healy13_PRE.88.042118,Healy12-PRL.109.170602}. 
Kim et al. \cite{JMKIM} studied the minimum energy distribution of directed polymer in random potential 
with Gaussian distribution up to $(3+1)$ dimensions and obtained  non-Gaussian distribution in those dimensions. Halpin-Healy and 
Takeuchi \cite{Healy-Takeuchi15}, throughly explored the statistics of the higher dimensional DPRM that yield non-zero skewness and kurtosis 
values. Alves et al. \cite{Alves-PRE-90-020103-2014} studied the  higher  dimensional KPZ height distributions via the RSOS model 
for flat initial condition and found the distributions to be non-Gaussian.

There have been a large number of numerical works on the $(2+1)$-dimensional KPZ type growth. For instance, the study of 
the RSOS model by Kim and Kosterlitz \cite{jmkim_prl_62_2289_89} leads to estimation of the scaling exponents that agree 
with their proposed relations $z=2(d+2)/(d+3)$ and $\beta=1/(d+2)$ for spatial dimensions $d\leq4$. 
Kondev et al. \cite{Kondev00} developed an approach wherein properties of scaling of
loops of constant height are analyzed to conclude upon geometrical and roughness exponents. They obtained the roughness 
exponent $\chi=0.38 \pm 0.08$ via nonlinear estimation. Quite a few growth models having a great deal of diversity, all of which belonging to the $(2 + 1)$-dimensional 
KPZ universality class, have been studied \cite{Healy12-PRL.109.170602} for  the morphology and statistics in the transient regime. 
The studied RSOS model, Euler integration of the KPZ equation and the mapping of the KPZ equation to a driven dimer model lead to 
roughness exponent $\chi = 0.383$, $0.388$ and $0.375$, respectively. On the other hand, for 
the stationary state, $\beta=0.241\pm0.001$ \cite{Healy13_PRE.88.042118} and thereby via the well known KPZ scaling 
relation $\chi(1+1/\beta)=2$ the roughness exponent is obtained as $\chi=0.387$--$0.390$. Kelling and Odor 
\cite{Kelling-Odor11_PRE.84.061150} performed a simulation considering a huge size up to $2^{17} \times 2^{17}$ 
and estimated the scaling exponents $\chi=0.393\pm0.004$ and $\beta=0.2415$ where the growth exponent of the simulation
is higher than $\beta=0.221$ \cite{Ghaisas06_PRE.73.022601} and $\beta=0.229$ \cite{PhysRevE.69.021610}.
Considering a potts-Spin representation via a multisite-coding with $11520^2$ sites, Forrest and Tang \cite{Forrest-Tang90_PRL.64.1405} 
estimated $\chi=0.385\pm 0.005$. An effort via a Monte-Carlo simulation of the hypercube-stacking model of 
Tang et al.\ \cite{Tang-Forrest-Wolf-PRA.45.7162} yields the growth exponent $\beta=0.240 \pm 0.001$. 
A numerically discretized RSOS model, studied by Marinari et al.\ \cite{Marinari00_JPA} by means of multi-surface coding, 
yields $\chi=0.393 \pm0.003$ and $\beta=0.244\pm 0.003$. Odor et al. \cite{Odor-PRE-81-031112-2010} found 
$\chi= 0.395\pm 0.005$ by mapping the driven lattice gases of $d$-dimers model onto the KPZ problem.

Theoretical calculation of $\chi$ in two and higher  dimensions has been a challenging work. Analytical approaches such as the
perturbative RG \cite{KPZ86,KPZ89,Nattermann92_PRA.45.7156,frey_pre_50_1024_94} and nonperturbative approaches, such as mode coupling 
\cite{Beijeren_PRL_54_2026,Bouchaud93_PRE.47.R1455,Frey96_PRE.53.4424,Tu_PhysRevLett.73.3109.1994} and self-consistent expansion 
\cite{Schwartz_Edwards_1992} are incapable of giving any conclusive scaling exponents as well as universality in $d=2$ dimension. 
L\"assig \cite{Lassig98_PRL.80.2366} employed an operator product expansion and obtained $\chi=2/5$ and $z=8/5$. A mode coupling 
calculation of Colaiori and Moore \cite{Colaiori_PRL_86_3946} suggested the dynamic exponent $z=1.62$ and roughness exponent 
$\chi=0.38$. A nonperturbative field theoretic RG has been employed by Kloss \cite{Kloss12_PRE.86.051124} in the stationary 
state and obtained roughness exponent $\chi \simeq 0.373$ via amplitude ratio of temporal and spatial correlation 
\cite{Kloss14_PRE.89.022108}. 

There have been experiments that mimic the $(2+1)$-dimensional KPZ scaling and the distribution of height fluctuations. 
Growth of oligmer thin film due to vapor deposition on a silicon substrate \cite{Palasantzas2002357} yields the roughness and 
growth exponents as $\chi=0.45\pm 0.04$ and $\beta=0.28\pm 0.05$. For the same system, the measured value of skewness $S=0.23$ 
\cite{Tsamouras01-APL} in the transient regime indicates that the growth is in the $(2+1)$-dimensional KPZ universality class. 
It is interesting to note that Almeida et al.\ \cite{Almeida14-PRB.89.045309} studied the height fluctuations on a polycrystalline
CdTe/Si(100) sample and measured $\beta=0.27 \pm 0.04$.

From the knowledge of geometry dependent subclasses in $(1+1)$ dimensions, it is well known that the scaling exponents
are not sufficient to understand the KPZ universality class. For the identification of the universality class, information about the 
whole distribution function is essential.  Moreover, it has been suggested that the measurements of moments are more stable and accurate 
\cite{Meakin93} than the scaling exponents.

Marinari et al. \cite{Marinari00_JPA} estimated higher order moments through multi-surface coding in different 
dimensions. From their reported moments in 2D, the skewness and kurtosis can be calculated as $|S|\sim0.266$ and 
$Q \sim 0.121$, respectively. Recently, two authors of the same group, Pagnani and Parisi \cite{Pagnani-Parisi-PRE-2015} 
refined the study of $(2+1)$-dimensional KPZ-type growth in the steady state 
and estimated two sets of best-fit results, namely, FIT-I and FIT-II for roughness exponent, skewness and kurtosis values. 
They found roughness exponent $\chi=0.3893 \pm0.0006$ (FIT-I), $\chi=0.3869\pm0.0004$ (FIT-II), skewness 
$|S|=0.2669\pm0.0004$ (FIT-I), $|S|=0.2657\pm0.0004$ (FIT-II) and kurtosis $Q=0.146\pm0.002$(FIT-I), $Q=0.145\pm 0.001$ (FIT-II).
Chin and den Nijs \cite{Chin-Marcel99PRE.59.2633} have performed a numerical study of the $(2+1)$-dimensional KPZ equation 
in the stationary state and obtained the roughness exponent $\chi \approx 0.38$ considering finite size scaling. 
They concluded that the third moment is more stable and more sensitive (than the roughness exponent) 
so that it is more suitable to determine and verify the universality class. They found skewness $|S|=0.27$ and excess kurtosis $Q=0.15$ 
which are the same as those in Kim-Kosterlitz (KK) and BCSOS models, thus identifying them to belong to the universality class of the 
$(2+1)$-dimensional KPZ dynamics.

Halpin-Healy \cite{Healy13_PRE.88.042118} has reported the value of average skewness ($S=0.244$) and kurtosis ($Q=0.177$)
for three models namely, RSOS, $g5_1$ DPRM and KPZ Euler. In the literature, the roughness and dynamic exponents ($\chi\approx0.39$, $z\approx1.61$) 
have been estimated from a considerable amount of numerical effort. For the purpose of calculating the skewness and kurtosis in $(2+1)$ dimensions,
 we take $\chi=9/23$ ( and $z=37/23$, satisfying $\chi+z=2$) as the sole input. The advantage of taking $\chi$ as a rational number is to
 avoid uncontrollable truncation errors in the subsequent exponents occuring in the calculations. We thus write the renormalized surface tension and 
 noise amplitude as
\beq
\nu(k)=A\ k^{-9/23}
\label{nuk}
\eeq
and 
\beq
D(k)= B \ k^{-27/23}
\label{Dk}
\eeq
where $A$ and $B$ are scale independent constants. The noise correlation in the Fourier space is written as 
\beq
\langle\eta(\vk,\omega)\eta(\vk',\omega')\rangle = 2D_0 (2\pi)^d\delta^d(\vk+\vk')(2\pi)\delta(\omega+\omega').
\eeq

Although the scaling exponents and universality class of $(1+1)$-dimensional KPZ have been studied extensively, there are few numerical 
estimations of moments in the $(2+1)$-dimensional case, whereas analytical treatments are extremely rare. In this paper, we calculate the 
higher order statistical moments of height fluctuation of the $(2+1)$-dimensional flat KPZ equation in the stationary state. 
This is achieved by calculating the cumulants up to the fourth order by employing a perturbation scheme to obtain 
the connected loop diagrams.

This paper is organized in the following way. Section II and Section III present the calculations of the third and fourth cumulants, respectively. 
In Section IV, we calculate the second cumulant. Skewness and kurtosis values are obtained from the calculated cumulants  in Section V. Finally discussions 
and conclusions are given in Section VI.

\section{The Third Cumulant}
Fourier transform of the KPZ equation (Eq. \ref{eq-kpz}) is written as  
\beq
(-i \omega+\nu_0 k^2) h(\vk,\omega)=\eta(\vk,\omega) -\frac{\lambda_0}{2} 
\int\!\!\int\frac{d^d\vq d\Omega}{(2\pi)^{d+1}}[\vq\cdot(\vk-\vq)]h(\vq,\Omega)h(\vk-\vq,\omega-\Omega),
\label{KPZFT}
\eeq
which will be used for perturbation calculations of cumulants.

The third cumuant $\langle h^3(\vx,t)\rangle_c$ can be expressed in the Fourier space as
\begin{eqnarray}
\lefteqn{
W_3=\langle h^3(\vx,t)\rangle_c =
\int\frac{d^dk_1\,d\omega_1}{(2\pi)^{d+1}}
\int\frac{d^dk_2\,d\omega_2}{(2\pi)^{d+1}}
\int\frac{d^dk_3\ d\omega_3}{(2\pi)^{d+1}} \nonumber}\\
&& \langle h(\mathbf{k}_1,\omega_1)\,h(\mathbf{k}_2,\omega_2)\,h(\mathbf{k}_3,\omega_{3})\rangle_c
\,\,e^{i(\mathbf{k}_1+\mathbf{k}_2+\mathbf{k}_3) \cdot \vx}
\,e^{-i(\omega_1+\omega_2+\omega_3)t}
\label{h^3}
\end{eqnarray}

\begin{figure}[ht!]
\begin{center}
\includegraphics[width=5.5cm]{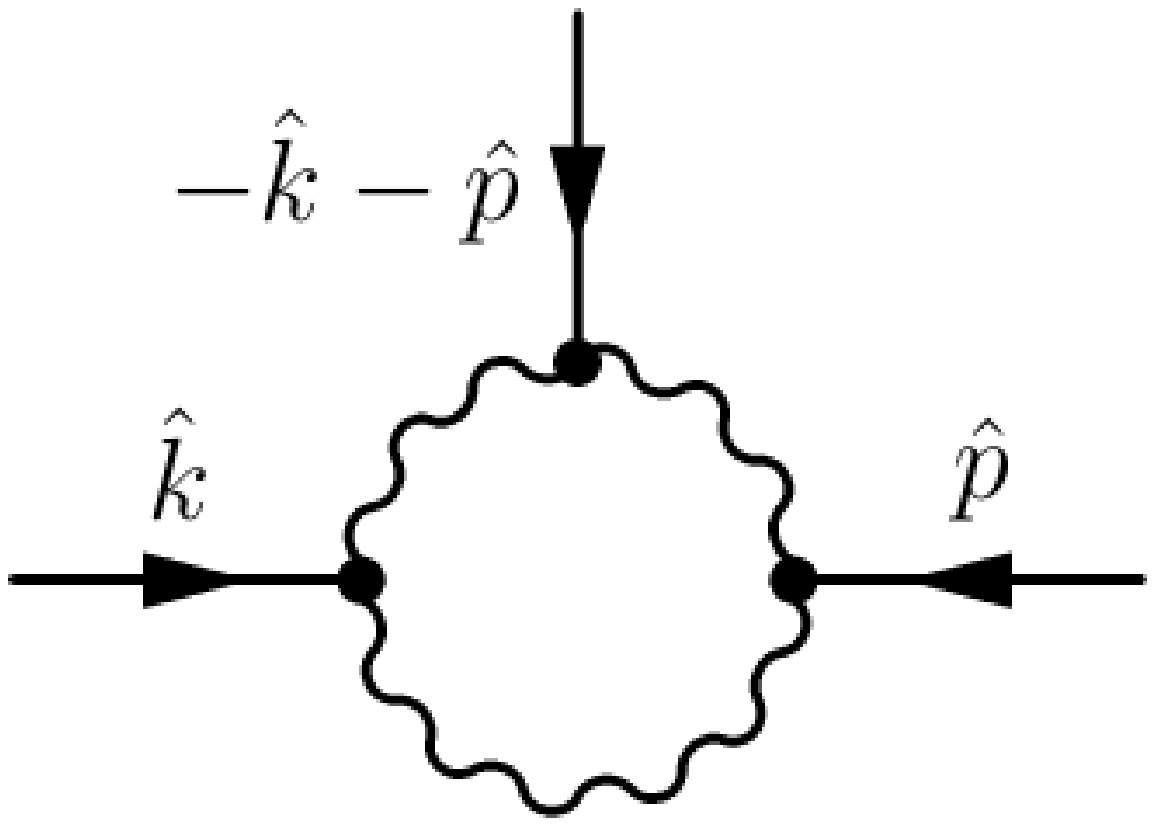}
\end{center}
\caption{Feynman diagram corresponding to the third cumulant where a wiggly line represents correlation and 
solid line response.}
\end{figure}

The third cumulant in Eq. \ref{h^3} is constructed by using Eq. \ref{KPZFT} in a perturbative frame-work. 
We follow the diagrammatic approach and obtain a one-loop diagram which contributes to $\langle h^3(\vx,t)\rangle_c$ as shown in Fig.~1.
Consequently, $\langle h^3(\vx,t)\rangle_c$ is written as  
\begin{equation}
\langle h^3(\vx,t)\rangle_c=\int \frac{d^{d+1}\hat{k}_1}{(2\pi)^{d+1}} \int
\frac{d^{d+1}\hat{k}_2}{(2\pi)^{d+1}} \, G(\hat{k}_1) \, G(\hat{k}_2)
\, L_3(\hat{k}_1;\hat{k}_2) \, G(-\hat{k}_1-\hat{k}_2)
\label{eq-L3-full}
\end{equation}
 where $L_3(\hat{k}_1;\hat{k}_2)$ indicates the amputated loop (excluding the external legs) and $\hat{k}_1$ stands for $(\vk_1, \omega_1)$. 
 We first consider bare value of the loop integral \cite{Singha-Nandy14_PRE.90.062402} which is expressed as 
\begin{eqnarray}
\lefteqn{
L^{(0)}_{3}(\vk_1,\omega_1;\vk_2,\omega_2)= 8 \left(\frac{-\lambda_0}{2} \right)^3 \int
\frac{d^d q\, d\Omega}{(2\pi)^{d+1}}\,\, [(\vq-\vk_1)\cdot(\vk_2+\vk_1-\vq)]
\,[\vq \cdot (\vk_1-\vq)] \nonumber }\\
&& [-\vq \cdot(\vq-\vk_2-\vk_1)] \, Q_0(\vq,\Omega) \,
Q_0(\vk_1-\vq,\omega_1-\Omega)\, Q_0(\vk_1+\vk_2-\vq,\omega_1+\omega_2-\Omega).
\label{eq-loop3}
\end{eqnarray}
Frequency and momentum integrations are performed in Eq. \ref{eq-loop3} in the limit of zero external momenta and 
frequencies. This gives the leading order contribution in an expansion when the external momenta and 
frequencies are small with respect to 
internal ones in the loop integral. The momentum is integrated in the thin shell $\Lambda_0 e^{-r} \leq q \leq \Lambda_0$, yielding
\beq
L^<_3(r)=\frac{3}{2} K_d \frac{\lambda^3_0 D^3_0}{\nu^5_0 \Lambda^{4-d}_0}
\frac{e^{(4-d)r}-1}{4-d}.
\eeq
where $K_d=S_d/(2\pi)^d$ with $S_d$ the surface area of unit sphere embedded in a $d$-dimensional space.
Assuming that shell elimination is performed in recursive 
steps \cite{yakhot_j_s_comput_1_3_86}, we obtain a differential equation for the scale dependent loop as 
\beq
\frac{dL_3}{dr}=\frac{3}{2} K_d \frac{\lambda^3_0 D^3(r) }{\nu^5(r) \Lambda^{4-d}(r)}
\label{DL3r}
\eeq
where $\Lambda(r)=\Lambda_0 e^{-r}$.
Using the scaling relations Eq. \ref{nuk} and \ref{Dk}, and identifying $k$ as $\Lambda_0 e^{-r}$, we integrate Eq. \ref{DL3r}
over $r$, and obtain
\beq
L_3(r)=\lambda^3_0 \frac{69}{328 \pi} \left(\frac{ B^3}{A^5}\right)  \Lambda^{-82/23}_0  e^{82r/23}
\eeq
for $d=2$. Since $L_3(\hat{k}_1;\hat{k}_2)$ appearing in  Eq. \ref{eq-L3-full} 
represents the (renormalized) loop diagram, its value is determined by the independent momenta $\hat{k}_1$ and $\hat{k}_2$ 
that flow along two internal lines belonging to the loop. Moreover, $L_3(\hat{k}_1;\hat{k}_2)$ should be
symmetric with respect to interchange of momenta $\hat{k}_1$ and $\hat{k}_2$ because the right hand expression in 
Eq. (\ref{eq-L3-full}) is expected to be symmetric with respect to the same momentum exchange.
Consequently, we construct the momentum dependence in $L_3(\hat{k}_1;\hat{k}_2)$ by considering 
$\Lambda_0 e^{-r}$ in Eq. (\ref{eq-L3-full})  as $k_1^{1/2} k_2^{1/2}$. To obtain the dependence on the corresponding external 
frequencies $\omega_1$ and $\omega_2$, we identify $(\Lambda_0 e^{-r})^{-41/23}$ as $k_i^{-41/23} f_1\left(\frac{\omega_i}{k_i^z}\right)$ 
where $f_1(.)$ is a dimensionless scaling function given by
\beq
k_i^{-41/23} f_1\left(\frac{\omega_i}{k_i^z}\right)=k_i^{143/23} \nu^4(k_1) |G(k_i,\omega_i)|^4
\label{dynamic-scaling}
\eeq
with $i=1,2$. We thus write
\beq
L_3(\hat{k}_1;\hat{k}_2)=\lambda^3_0 \frac{69}{328 \pi} \left(\frac{B^3}{A^5}\right) k_1^{143/23} k_2^{143/23}  \nu^4(k_1) \nu^4(k_2)
|G(\vk_1,\omega_1)|^4 |G(\vk_2,\omega_2)|^4.
\label{L3final}
\eeq
Using the expression from Eq. \ref{L3final} in Eq. \ref{eq-L3-full}, we obtain 
$$
\langle h^3(\vx,t)\rangle_c=\lambda^3_0 \frac{69}{328 \pi} A^3 B^3 \int \frac{d^{d+1}\hat{k}_1}{(2\pi)^{d+1}} \int 
\frac{d^{d+1}\hat{k}_2}{(2\pi)^{d+1}}  k_1^{107/23} k_2^{107/23}  |G(\hat{k}_1)|^4  |G(\hat{k}_2)|^4  G(\hat{k}_1) 
G(\hat{k}_2)  $$ \beq  G(-\hat{k}_1-\hat{k}_2)\eeq
We perform the frequency integrations over $\omega$ and $\omega'$, leading to 
\beq
\langle h^3(\vx,t)\rangle_c=\lambda^3_0 \left(\frac{B}{A^2}\right)^3  \frac{69}{328 \pi} \int \frac{d^{2}k_1}{(2\pi)^{2}} 
\int \frac{d^{2}k_2}{(2\pi)^{2}} F(\vk_1,\vk_2)
\label{W3intefkp}
\eeq
The algebric form of the function $F(\vk_1,\vk_2)$ is given in Appendix.
We perform the integrations in Eq. \ref{W3intefkp} in cartesian coordinates. The function $F(\vk_1,\vk_2)$ is symmetric with respect to 
interchange of $\vk_1$ and $\vk_2$. Consequently, we can write 

\beq
I= \int d^2 k_1 \int d^2 k_2 F(\vk_1,\vk_2) = 4 [ I_1(\mu) +2 I_2(\mu) +I_3(\mu) ]
\label{int-I}
\eeq
where
\beq
I_1(\mu)=\int^{\infty}_{0} dk_{1x}\int^{\infty}_{0} dk_{2x} \int^{\infty}_{\mu} dk_{1y} \int^{\infty}_{\mu} dk_{2y} F(k_{1x},k_{1y},k_{2x},k_{2y}),
\eeq
\beq
I_2(\mu)=\int^{\infty}_{0} dk_{1x}\int^{\infty}_{0} dk_{2x} \int^{\infty}_{\mu} dk_{1y} \int^{\infty}_{\mu} dk_{2y} F(k_{1x},-k_{1y},k_{2x},k_{2y}),
\eeq
and 
\beq
I_3(\mu)=\int^{\infty}_{0} dk_{1x} \int^{\infty}_{0} dk_{2y} \int^{\infty}_{\mu} dk_{2x} \int^{\infty}_{\mu} dk_{1y}  F(-k_{1x},-k_{1y},k_{2x},k_{2y})
\eeq
where we have introduced an infrared cutoff $\mu$ because these integrals diverges at the lower limit.
We perform numerical integrations of these functions over $k_{1x}$, $k_{1y}$, $k_{2x}$ and $k_{2y}$, leading to the values
\begin{eqnarray}
I_1(\mu) =0.032196 \ \mu^{-27/23}, \\
I_2(\mu) =0.062963 \ \mu^{-27/23}, \\
I_3(\mu)=0.0043277 \ \mu^{-27/23},
\end{eqnarray}
for very small values of $\mu$ close to zero.
The value of the third cumulant coming from Eq. \ref{W3intefkp}, in terms of these integration values, is given by 
\beq
\langle h^3(\vx,t)\rangle_c= \left(\frac{\lambda_0B}{A^2}\right)^3  \frac{69}{328 \pi} \frac{1}{4 \pi^4} \left[I_1(\mu)+2 I_2(\mu)+I_3(\mu) \right].
\label{W3-I123}
\eeq
\section{The Fourth Cumulant}
The fourth order cumulant, written in Fourier space, assumes the form  
\begin{eqnarray}
\langle h^4(\vx,t) \rangle_c &=& \int \frac{d^{d+1} \hat{k}_1}{(2\pi)^{d+1}}\int
\frac{d^{d+1}\hat{k}_2}{(2\pi)^{d+1}} \!\int
\frac{d^{d+1} \hat{k}_3}{(2\pi)^{d+1}}
\!\int \frac{d^{d+1}\hat{k}_4}{(2\pi)^{d+1}}
\nonumber \\
 &&
 \langle h(\hat{k}_1) h(\hat{k}_2) h(\hat{k}_3) h(\hat{k}_4)\rangle_c 
\,e^{i(\hat{k}_1+\hat{k}_2+\hat{k}_3+\hat{k}_4)\cdot \hat{x}}.
\label{h4ini}
\end{eqnarray}
Following the diagrammatic approach, we obtain a connected loop diagram for the fourth order cumulant 
in Fourier space occurring in the integrand. The corresponding loop 
diagram is shown in Fig. 2, which suggests the expression
\begin{figure}[h!t]
\begin{center}
\includegraphics[width=4.3cm]{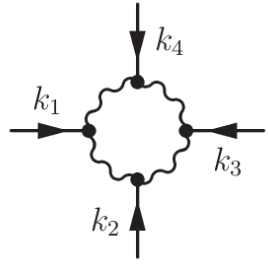}
\end{center}
\caption{Feynman diagram corresponding to the fourth cumulant where a wiggly line represents correlation and 
solid line response.}
\end{figure}

\begin{eqnarray}
 \langle h^4(\vx,t)\rangle_c &=& \int\frac{d^{d+1}\hat{k}_1}{(2\pi)^{d+1}}\int
\frac{d^{d+1} \hat{k}_2}{(2\pi)^{d+1}} \!\int
\frac{d^{d+1} \hat{k}_3}{(2\pi)^{d+1}} \,G(\hat{k}_1)\,G(\hat{k}_2)\, G(\hat{k}_3)
\nonumber\\
&& L_4(\hat{k}_1; \hat{k}_2; \hat{k}_3) \, G(-\hat{k}_1-\hat{k}_2-\hat{k}_3),
 \label{h4L4}
\end{eqnarray}
where $L_4(\hat{k}_1; \hat{k}_2; \hat{k}_3)$ is the contribution coming from the renormalized amputated loop (without the external legs)
and $G(\hat{k}_i)$ are the renormalized propagators.

 The unrenormalized expression for the amputated loop, in $(d+1)$ dimensions, corresponding to the fourth order cumulant is 
expressed as \cite{Singha-Nandy15_jstatmech}

 \begin{eqnarray}
L^{(0)}_4(\hat{k}_1; \hat{k}_2; \hat{k}_3) &=&
16 \left(-\frac{\lambda_0}{2}\right)^4 (2D_0)^4
\int \frac{d^{d+1}\hat{q}}{(2\pi)^{d+1}} [\vq
\cdot(\vq-\vko)][\vq\cdot(\vkt+\vq)]
\nonumber\\
&&
[(\vq+\vkt)\cdot(\vkth+\vkt+\vq)]
[(\vq-\vko)\cdot(\vq+\vkt+\vkth)] G_0(\hat{q}) \nonumber \\
&& G_0(-\hat{q}+\hat{k}_1) G_0(-\hat{q}) G_0(\hat{q}+\hat
{k}_2) 
G_0(-\hat{q}-\hat{k}_2) G_0(\hat{q}+\hat{k}_3+\hat{k}_2)
\nonumber \\
&&
G_0(\hat{q}-\hat{k}_1)G_0(-\hat{q}-\hat{k}_2-\hat{k}_3),
\label{L4}
\end{eqnarray}
 where the suffix $0$ signifies unrenormalized quantities. Carrying out the frequency and momentum integration in Eq. \ref{L4}, 
 we obtain the following expression  on elimination of modes from the shell $\Lambda(r) \leq q \leq \Lambda_0$.
\beq
L_4^{<}(r)= \frac{5}{2} K_d \frac{\lambda^4_0 D_0^4}{\nu_0^7 \Lambda^{6-d}_0} 
\, \frac{e^{(6-d)r}-1}{d-6}
\eeq
Assuming that shell elimination is performed in recursive 
steps, we obtain 
\beq
\frac{dL_4}{dr} =\frac{5}{2} K_d \frac{\lambda^4_0 D^4(r) }{\nu^7(r) \Lambda^{6-d}(r)}
\label{DifL4}
\eeq
Employing Eq. \ref{nuk} and \ref{Dk} with $k$ identified as $\Lambda_0 e^{-r}$ and integrating Eq. \ref{DifL4} over $r$ for $d=2$ 
substrate dimensions, we obtain 
\beq
L_4(r)=\frac{115}{548 \pi} \left[\frac{\lambda^4_0 D^4(r)}{\nu^7(r) \Lambda^{4}(r)}\right] .
\label{L4-r-finl}
\eeq
Since the loop $L_4(\hat{k}_1; \hat{k}_2;\hat{k}_3)$ depends on the three external momenta $k_1$, $k_2$ and $k_3$,
the expression for $L_4(\hat{k}_1;\hat{k}_2;\hat{k}_3)$ 
appearing in Eq. \ref{h4L4} is expected to be symmetric with respect to 
interchange of $\hat{k_1}$, $\hat{k_2}$ and $\hat{k_3}$. Thus the momentum dependence is constructed by considering 
$\Lambda_0 e^{-r}$ in Eq. \ref{L4-r-finl}  as $k^{1/3}_1 k^{1/3}_2 k^{1/3}_3$. To obtain the dependence on external 
frequencies $\omega_1$, $\omega_2$ and $\omega_3$, we identify 
$(\Lambda_0 e^{-r})^{-137/69}$ as $k_j^{-137/69} f_2\left(\frac{\omega_j}{k_j^z}\right)$ where $f_2(.)$ is a 
dimensionless scaling function. The form of the scaling function is introduced as
\beq
k_j^{-137/69} f_2\left(\frac{\omega_j}{k_{j}^z}\right)= k_j^{415/69} \nu^4(k_j) |G(k_j,\omega_j)|^4
\eeq
where $\hat k_j$ represents $\hat{k}_1$ or $\hat{k}_2$ or $\hat{k}_3$. Employing this scaling relation, the renormalized loop 
in Fig.\ 3 assumes the form
\beq
L_4(\hat{k}_1;\hat{k}_2;\hat{k}_3) =\lambda^4_0  \left(\frac{B^4}{A^7}\right) \frac{115}{548 \pi}
\prod^{3}_{j=1} k_j^{415/69} \nu^4(k_j) |G(\hat{k}_j)|^4
\label{L4final}
\eeq
Using the expression from Eq. \ref{L4final} in Eq. \ref{h4L4}, we obtain 
\beq
\begin{array}{rcl}
\langle h^4(\vx,t)\rangle_c & = & \displaystyle\lambda^4_0 A^5 B^4 \frac{115}{548 \pi}  \int \frac{d^{d+1}\hat{k}_1}{(2\pi)^{d+1}} 
\int \frac{d^{d+1}\hat{k}_2}{(2\pi)^{d+1}} \int \frac{d^{d+1}\hat{k}_3}{(2\pi)^{d+1}} G(-\hat{k}_1-\hat{k}_2-\hat{k}_3) \\
&& \displaystyle\prod^{3}_{j=1} k_j^{307/69} |G(\hat{k}_j)|^4 G(\hat{k}_j)
\end{array}
\eeq
where the response function involves the renormalized surface tension $\nu(\vk)$. 
Carrying out the frequency integrations yields
\begin{equation}
\langle h^4(\vx,t)\rangle_c =  \left(\frac{\lambda_0 B}{A^2}\right)^4 \frac{115}{548 \pi} \int \frac{d^{2}k_1}{(2\pi)^{2}}
 \int \frac{d^{2} k_2}{(2\pi)^{2}} \int \frac{d^{2} k_3}{(2\pi)^{2}} T(\vk_1,\vk_2,\vk_3)
\label{h4-Tkpq}
\end{equation}
where the form of $T(\vk_1,\vk_2,\vk_3)$ is given in Appendix. For brevity in notations, henceforth we rename the 
momenta $\vk_1$, $\vk_2$ and $\vk_3$ as $\vk$, $\vp$ and $\vq$.
Subsequently, we express the integrations in cartesian coordinates, so that
\begin{eqnarray}
 \langle h^4(\vx,t)\rangle_c =&& \frac{\lambda^4_0}{(2\pi)^6} \frac{B^4}{A^8} \frac{115}{548 \pi} [\int^{\infty}_{-\infty} dk_{x} 
\int^{\infty}_{-\infty} dk_{y}\int^{\infty}_{-\infty} dp_{x}\int^{\infty}_{-\infty} dp_{y}   \\
&& \int^{\infty}_{-\infty} dq_{x} \int^{\infty}_{-\infty} dq_{y}T(k_{x},k_{y};p_{x},p_{y};q_{x},q_{y}) ] \nonumber
\label{W4intf}
\end{eqnarray}
Due to the symmetry of the integrand in momentum variables, we can break the function as
 \begin{eqnarray}
 &&\langle h^4(\vx,t)\rangle_c= \frac{4}{(2\pi)^6} \left(\frac{\lambda_0 B}{A^2}\right)^4 \frac{115}{548 \pi} 
 \int^{\infty}_{0} dk_{x} \int^{\infty}_{0} dk_{y} \int^{\infty}_{0} dp_{x}
 \int^{\infty}_{0} dp_{y}  \int^{\infty}_{0} dq_{x} \int^{\infty}_{0} dq_{y} \nonumber \\
 &&   \{T(k_{x},k_{y};p_{x},p_{y};q_{x},q_{y}) +6  T(k_{x},-k_{y};p_{x},p_{y};q_{x},q_{y})+9 T(-k_{x},-k_{y};p_{x},p_{y};q_{x},q_{y})\}
\end{eqnarray}
The momentum dependence causes the integrations to diverge in the infrared limit. 
Therefore, we set $\mu$ as the lower cutoff of the integrations  
and express the integration in terms of  
\beq
J_1(\mu)=\int^{\infty}_{0} dk_{x}\int^{\infty}_{0} dk_{y}\int^{\infty}_{0} dp_{x}\int^{\infty}_{\mu} dp_{y}  
 \int^{\infty}_{\mu} dq_{x}\int^{\infty}_{\mu} dq_{y}  T(k_{x},k_{y};p_{x},p_{y};q_{x},q_{y}), 
\eeq
\beq
J_2(\mu)=\int^{\infty}_{0} dk_{x}\int^{\infty}_{0} dk_{y}\int^{\infty}_{0} dp_{x}\int^{\infty}_{\mu} dp_{y}  
 \int^{\infty}_{\mu} dq_{x}\int^{\infty}_{\mu} dq_{y} T(k_{x},-k_{y};p_{x},p_{y};q_{x},q_{y}) 
\eeq
and 
\beq
J_3(\mu)=\int^{\infty}_{0} dk_{x}\int^{\infty}_{0} dk_{y}\int^{\infty}_{0} dp_{x}\int^{\infty}_{\mu} dp_{y}  
 \int^{\infty}_{\mu} dq_{x}\int^{\infty}_{\mu} dq_{y} T(-k_{x},-k_{y};p_{x},p_{y};q_{x},q_{y}) 
\eeq
so that
\begin{equation}
\langle h^4(\vx,t)\rangle_c= \frac{4}{(2\pi)^6} \left(\frac{\lambda_0 B}{A^2}\right)^4 \frac{115}{548 \pi} 
\left[J_1(\mu)+6 J_2(\mu)+9 J_3(\mu)\right].
\label{h4-J123}
\end{equation}
Performing the integrations numerically, we obtain 
\beq
J_1(\mu)= 0.0069369 \mu^{-36/23},
\eeq
\beq
J_2(\mu)= 0.0000151\mu^{-36/23},
\eeq
and
\beq
J_3(\mu)=0.0219628 \mu^{-36/23}.
\eeq
We shall employ these numerical values in the next Section while calculating the value of kurtosis.

\section{The Second Cumulant}
In order to calculate the skewness and kurtosis, we need in addition to the third and fourth, the second cumulant. 
The first moment, $\langle h(\vx,t) \rangle$, is zero in the steady state, where the angular brackets denote as ensemble average. 
The second cumulant is expressed in the Fourier space as
\begin{equation}
W_2= \langle h^2(\vx,t)\rangle_c =\int \frac{d^dk d \omega}{(2\pi)^{d+1}} \int \frac{d^dk' d\omega'}{(2\pi)^{d+1}} \langle
 h(\vk,\omega)h(\vk',\omega')\rangle_c  e^{i(\vk+\vk') \cdot \vx} e^{-i(\omega+\omega')t}.
\label{2Dh^2}
\end{equation}
where the height-height correlation is given by
\beq
\langle h(\vk,\omega)  h(\vk',\omega')\rangle= Q(\vk,\omega) (2\pi)^d\delta^d(\vk+\vk')(2\pi)\delta(\omega+\omega') 
\label{corre-KPZ2D}
\eeq
with $Q(\vk,\omega)$ the renormalized correlation. Using Eq. (\ref{corre-KPZ2D}) in Eq. (\ref{2Dh^2}) leads to 
\beq
\langle h^2(\vx,t)\rangle_c= \int \frac{d^dk\ d\omega}{(2\pi)^{d+1}} Q(\vk,\omega).
\label{W2FT}
\eeq
The correlation, $Q(\vk,\omega)$, can be obtained perturbatively using Eqs. (\ref{KPZFT}) and (\ref{corre-KPZ2D}). 
To obtain the leading order contribution to the second cumulant, we consider the one loop diagram shown in Fig.3, that
satisfies the expression
\beq
Q(\vk,\omega)=G(\vk,\omega) L_2(\vk,\omega)G(-\vk,-\omega)
\label{corre-Q}
\eeq

\begin{figure}[h!t]
\begin{center}
\includegraphics[width=4.6cm]{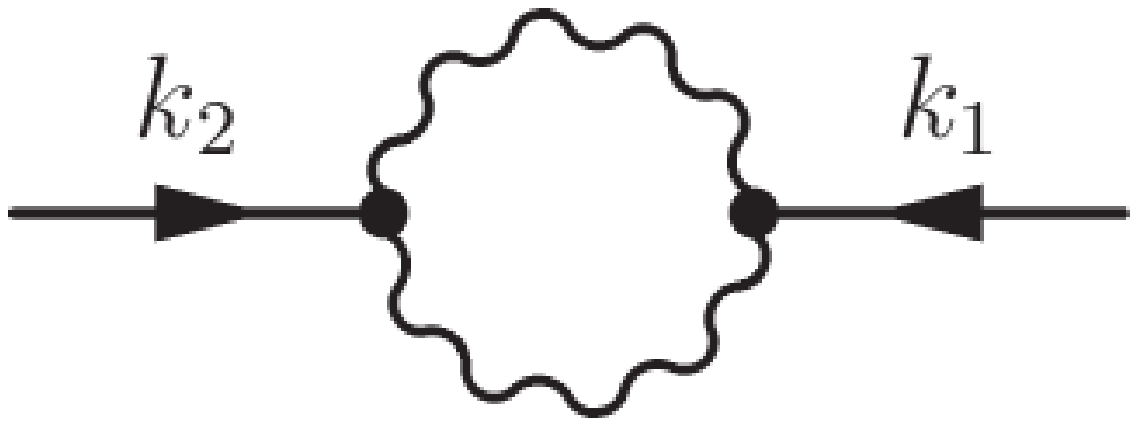}
\end{center}
\caption{Feynman diagram corresponding to the second cumulant where a wiggly line represents correlation and 
solid line response.}
\end{figure}
Equation \ref{corre-Q} corresponds to the loop diagram in Fig.3.
The unrenormalized amputated loop (excluding the external legs) is expressed as 
\beq
L^{(0)}_2(\vk,\omega)= 2 \left(\frac{-\lambda_0}{2}\right)^2 \int^{\Lambda_0}_{\Lambda (r)} \frac{d^d q}{(2\pi)^{d}} 
\int^{\infty}_{-\infty} \frac{d \omega}{(2\pi)}[\vq \cdot(\vk-\vq)]^2 |G_0(\hat{\vq})|^2 |G_0(\hat{\vk}-\hat{\vq})|^2|(2D_0)^2
\eeq
that contains unrenormalized noise amplitude $D_0$ and unrenormalized response function $G_0(\vk,\omega)=[-i \omega+\nu_0 k^2]^{-1}$.
Carrying out the frequency integration in the whole range and the momentum integration in the shell 
$\Lambda_0 e^{-r} \leq q \leq \Lambda_0$, we obtain 
\beq
L^{<}_2(r)= \frac{\lambda_0^2 D^2_0 K_d}{2\nu_0^3} \left[\frac{\Lambda^{d-2}_0-\Lambda^{d-2}(r)}{(d-2)}\right].
\eeq
We construct a differential equation for $L_2$ with respect to the scale parameter $r$, 
\beq
\frac{dL_2}{dr}=K_d \frac{\lambda_0^2  D^2(r)}{2\nu^3(r) \Lambda^{2-d}(r)}
\eeq
where $\Lambda(r)=\Lambda_0 e^{-r}.$ Using the scale dependence from Eqs. \ref{nuk} and \ref{Dk}, and integrating over $r$ leads to
\beq
L_2(r)=\frac{23}{108 \pi}  \left(\frac{\lambda^2_0 B^2}{A^3}\right) (\Lambda_0 e^{-r})^{-27/23}
\eeq
for $d=2$. We consider the scale dependent parameter $(\Lambda_0 e^{-r})^{-27/23}$ as $ k^{-27/23} \ f\left(\frac{\omega}{k^z}\right)$, where 
$ f\left(.\right)$ dimensionless scaling function and $z$ is the dynamic exponent.
Thus $L_2(\vk;\omega)$ is expressed as
\beq
L_2(\vk;\omega)=\frac{23}{108 \pi} \left(\frac{\lambda^2_0 \ B^2}{A^3}\right) k^{-27/23} \ f\left(\frac{\omega}{k^z}\right).
\eeq
As before, we write
\beq
k^{-27/23} f\left(\frac{\omega}{k^z}\right)=k^{3861/943} \nu^{108/41}(k) |G(\vk,\omega)|^{108/41}
\eeq
so that $L_2(\vk;\omega)$ becomes 
\beq
L_2(\vk;\omega)=\frac{23}{108 \pi} \left(\frac{\lambda^2_0 \ B^2}{A^3}\right) k^{3861/943} \nu^{108/41}(k) \ |G(\vk,\omega)|^{108/41}
\label{L2-final}
\eeq
From Eqs. \ref{W2FT} and \ref{corre-Q}, we have
\beq
\langle h^2(\vx,t)\rangle_c=\int \frac{d^d k}{(2\pi)^d} \int \frac{d \omega}{(2\pi)} |G(\vk,\omega)|^2 L_2(\vk;\omega)
\eeq
where we substitute the expression for $L_2$ given by Eq. \ref{L2-final}, so that
\beq
\langle h^2(\vx,t)\rangle_c=\frac{23}{108 \pi} \left(\frac{\lambda^2_0 B^2}{A^{3}}\right) A^{108/41} \int \frac{d^2 k}{(2\pi)^2} k^{2889/943} 
\int \frac{d \omega}{(2\pi)} \frac{1}{[\omega^2+\nu^2(k) k^4]^{95/41}},
\eeq
where the surface tension $\nu(k)$ in the denominator is renormalized.
Performing the frequency integration and expressing the momentum integration in cartesian coordinates, we obtain
\beq
\langle h^2(\vx,t)\rangle_c=\frac{23}{108 \pi} \left(\frac{\lambda_0 B}{A^{2}}\right)^2 \left(\frac{\Gamma \left(\frac{149}{82} \right)}{2\sqrt{\pi} \ \Gamma \left(\frac{95}{41} \right)}\right)
\left[\frac{4}{4 \pi^2} \int^{\infty}_{0} dk_x \int^{\infty}_{0} dk_y \frac{1}{[k_x^2+k_y^2]^{32/23}}\right].
\label{W2-kx-ky}
\eeq

The factor of 4 appears because the integrand is an even function of $k_x$ and $k_y$. Moreover, the integration in the Eq. \ref{W2-kx-ky} is 
symmetric with respect to interchange of $k_x$ and $k_y$. We first do the momentum integration over $k_x$ with limits, $0 \leq k_x \leq \infty$
yielding

\beq
\langle h^2(\vx,t)\rangle_c=\frac{23}{108 \pi^3} \left(\frac{\lambda_0 B}{A^{2}}\right)^2  
\left(\frac{\Gamma \left(\frac{149}{82} \right)}{2\sqrt{\pi} \ \Gamma \left(\frac{95}{41} \right)}\right) 
\left(\frac{\sqrt{\pi}}{2} \frac{\Gamma \left(\frac{41}{46}\right)}{\Gamma \left(\frac{32}{23}\right)} \right)  
\left[\int^{\infty}_{\mu} dk_y k_y^{-41/23} \right]
\eeq

We have introduced an infrared cutoff $\mu$ as the integral on $k_y$ diverges at the lower limit. 
We thus obtain by integration
\beq
\langle h^2(\vx,t)\rangle_c=  [0.211105 \times 10^{-2}] \  \left(\frac{\lambda_0 B}{A^2} \right)^2 \frac{1}{\mu^{18/23}}
\label{W2-final}
\eeq
where we have used $\Gamma \left(\frac{149}{82} \right)=0.936026$, $\Gamma \left(\frac{95}{41} \right)=1.178818$, 
$\Gamma \left(\frac{41}{46}\right)=1.075746$ and $\Gamma \left(\frac{32}{23} \right)=0.887772$.
This is the second cumulant of the height fluctuations in the stationary state of the $2+1$ KPZ equation.

\section{Skewness and Kurtosis}
In this section, we calculate the skewness and kurtosis of the height fluctuations obeying the $(2+1)$-dimensional 
KPZ equation in the stationary state. For this purpose, we use the values of the second, third and fourth order 
moments evaluated above. The $n$ th moments is expressed as 
\beq
W_n=\left \langle [h(\vx,t)-\langle h\rangle]^n \right \rangle
\label{Wn}
\eeq
which is related to the system size as $W_n \sim L^{n \chi}$, where $L$ is the size of the system.
Substituting $n=3$ and $4$ in Eq. \ref{Wn}, we obtain the expressions
\beq
W_2= \langle h^2\rangle - \langle h\rangle^2,
\eeq
\beq
W_3=\langle h^3 \rangle -3 \langle h \rangle \langle h^2 \rangle + 2 \langle h \rangle^3 \nonumber \\
\eeq
\beq
W_4=\langle h^4 \rangle -4\langle h \rangle \langle h^3 \rangle + 6 \langle h^2 \rangle  \langle h \rangle^2 -3 \langle h^2 \rangle^2.
\label{W4}
\eeq
The moments and cumulants are related as
\begin{eqnarray}
\langle h \rangle=&&\langle h \rangle_c  \nonumber \\
\langle h^2 \rangle=&&\langle h^2 \rangle_c + \langle h \rangle^2_c \nonumber
\\
\langle h^3 \rangle=&&\langle h^3 \rangle_c +3 \langle h \rangle_c\langle h^2
\rangle_c +\langle h \rangle^3_c \nonumber \\
\langle h^4 \rangle=&&\langle h^4 \rangle_c +4\langle h \rangle_c \langle h^3
\rangle_c+3 \langle h^2 \rangle^2_c +6\langle h \rangle^2_c \langle h^2
\rangle_c + \langle h \rangle^4_c  \nonumber \\
\end{eqnarray}
These higher order moments determine the values of skewness and kurtosis.
Taking ensemble average on both sides of Eq. (6), we find that both terms on the right hand side vanish because the 
noise is Gaussian and the second term yields a Dirac delta function $\delta^d(\vk)$ upon using Eq. (45). 
Finiteness of the substrate (although it is assumed to be large) implies that $\langle h\rangle=0$. 
Thus, skewness and kurtosis may be expressed as   
\beq
S=\frac{\langle h^3 \rangle}{\langle h^2 \rangle^{3/2}}=\frac{\langle h^3 \rangle_c}{\langle h^2 \rangle_c^{3/2}}
\label{def-kurto}
\eeq
and
\beq
Q=\frac{\langle h^4 \rangle}{\langle h^2 \rangle^{2}}-3=\frac{\langle h^4 \rangle_c}{\langle h^2 \rangle_c^2}
\label{defh4}
\eeq

where the suffix $c$ indicates the cumulants that correspond to connected diagramms in the perturbative expansion. 
Skewness $S$ measures the asymmetry of the distribution function with respect to the Gaussian. 
A positive (negative) value of skewness is obtained when the distribution has a longer tail on the right (left) side.
The $Q$ value indicates sharpness with respect to the Gaussian distribution. A positive (negative) value of $Q$ 
signifies a sharper (flatter) distribution than the Gaussian.

We obtain the skewness and kurtosis employing the above definitions. Using the numerically evaluated integrations 
of $I_1(\mu)$, $I_2(\mu)$ and $I_3(\mu)$ in Eq. \ref{W3-I123}, we obtain
\beq
\langle h^3(\vx,t) \rangle_c=[0.027922 \times 10^{-3}] \left(\frac{\lambda_0B}{A^2}\right)^3 \mu^{-27/23}.
\label{W3-final}
\eeq

Substituting from  equations \ref{W2-final} and \ref{W3-final} in Eq. \ref{def-kurto}, we calculate the skewness 
of the $(2+1)$-dimensional KPZ height distribution as
\beq
S=\frac{\langle h^3(\vx,t)\rangle_c}{\langle h^2(\vx,t)\rangle_c^{3/2}}=\frac{0.027922}{(0.211105)^{3/2}}= 0.2879.
\label{skewness2D}
\eeq

Similarly, we substitute the results of the evaluated numerical integrations $J_1(\mu)$, $J_2(\mu)$ and $J_3(\mu)$ in Eq. \ref{h4-J123},
obtaining
\begin{equation}
\langle h^4(\vx,t)\rangle_c= [0.008889 \times 10^{-4}] \left(\frac{\lambda_0 B}{A^2}\right)^4 \mu^{-36/23}.
\label{W4-final}
\end{equation}

Using Eq. \ref{W4-final} and Eq. \ref{W2-final} in Eq. \ref{defh4} 
leads to the kurtosis value as
\begin{equation}
 Q=\frac{\langle h^4(\vx,t)\rangle_c}{\langle h^2(\vx,t)\rangle^2_c}=\frac{0.008889}{(0.211105)^2}=0.1995
\label{kurtosis2D}
\end{equation}

We note that these values for skewness and kurtosis are obtained for the $(2+1)$-dimensional KPZ dynamics corresponding to the stationary state.

\section{Discussion and Conclusion}

Our main motivation in this work comes from two facts. First, the growth of a surface on a 2D substrate is a 
commonly occurring phenomenon in Nature. Second, analytical methodologies to obtain the skewness and kurtosis 
values directly from the dynamical equation are unavailable in the existing literature. The obtained skewness 
$S=0.2879$ and kurtosis $Q=0.1995$ values are independent of model parameters ($\nu_0$, $D_0$, and $\lambda_0$), 
scaling coefficients ($A$ and $B$) and the momentum cutoffs ($\Lambda_0$ and $\mu$) in the calculations. 
The sole input to our calculations is the roughness exponent $\chi=9/23=0.391304$ which is a good approximation 
to high resolution numerical results.

In this context it may be noted that most analytical approaches have been unsuccessful to obtain the scaling exponents in $(2+1)$ dimensions, 
apart from the works of L\"assig \cite{Lassig98_PRL.80.2366}, Tu \cite{Tu_PhysRevLett.73.3109.1994}
Colaiori and Moore \cite{Colaiori_PRL_86_3946} and Kloss et al.\cite{Kloss14_PRE.89.022108}, 
as mentioned earlier. At the same time, a huge amount of numerical approaches suggest the roughness and dynamic  exponents to be 
$\chi \approx 0.39$ and $z \approx 1.61$, respectively. 

We employed perturbation theory directly on the KPZ equation to obtain expressions for 
$L^{<}_3(r)$, $L^{<}_4(r)$ and $L^{<}_2(r)$ that contain the bare parameters $\nu_0$ and $D_0$. Obtaining these 
expressions solely depend on the use of perturbation theory and they do not incorporate the renormalization
group in the conventional sense. We use these expressions for $L^{<}_3(r)$, $L^{<}_4(r)$, and $L^{<}_2(r)$ 
to obtain the flow equations for the renormalized quantities $L_3(r)$, $L_4(r)$, and $L_2(r)$ that involve 
the renormalized quantities $\nu(r)$ and $D(r)$. In this process, we are able to find explicit mathematical 
relations including the prefactors once the scaling laws for the effective surface tension $\nu(k)$
and the noise amplitude $D(k)$ are assumed (Eqs. (3) and (4)) in consistency 
with the scaling relation $\chi+z=2$. We incorporate frequency 
dependence of these loops by scaling functions that preserve their 
real valuedness and their correct zero frequency limits. This allows for the 
calculation of the corresponding cumulants that are found to depend  
on the infrared cutoff $\mu$ as $W_n \sim \mu^{-n \chi}$.  This is expected because the cumulants have the 
 semi-extensive property $W_n \sim L^{n \chi}$ in the stationary state where $L$ is the substrate size. 
Thus the infrared cutoff can be identified as $\mu \sim L^{-1}$. We finally obtain the skewness value 
$S=0.2879$ and the kurtosis value $Q=0.1995$, relevant to the case of $(2+1)$-dimensional KPZ growth in the 
stationary state.

It may be noted that it is not possible to incorporate the results of the standard renormalization group analyses
that do not yield a strong coupling fixed point and thereby providing no prediction 
for the value of $\chi$ in two dimensions.  
On the other hand, mode-coupling theories suggest that the upper critical dimension is $3.6$, $4$ or $\infty$ 
\cite{Doherty_PRL_94,Colaiori_PRL_86_3946,Tu_PhysRevLett.73.3109.1994}. Interestingly, a non-perturbative renormalization
group analysis \cite{Kloss12_PRE.86.051124} indicated the existence of a stable strong coupling fixed point for 
$d\leq 2$, whereas for $d>2$, there exist two basins of attraction containing a Gaussian fixed point and 
a strong coupling fixed point. The resulting roughness exponent was found to be $\chi=0.330(8)$ and
$\chi=0.373(1)$ (in two dimensions) in the leading and next to leading order approximations, respectively. 
The latter result agrees very well with the numerical estimation $\chi \approx 0.39$ that we have used in our calculations. We further note that the values for the amplitudes $A$ and $B$ (that determine the fixed point value $g^{*} \sim \lambda_0^2 B/A^3$ ) are not required in our calculations because they cancel out in the ratios determining $S$ and $Q$.           

It can be seen that the scalings of the renormalized quantities are $\nu(k) \sim k^{-\chi}$ and 
$D(k)\sim k^{-3 \chi+2-d}$ in $d$ dimensions. Consequently, the scalings for the loop functions turn out to be 
$L_3(k) = C_3 \, k^{-4\chi-2d+2}$, $L_4(k)=C_4 \, k^{-5\chi-3d+2}$ and $L_2(k)=C_2 \, k^{-3\chi-d+2}$. We expect these scaling relations to be correct for any  (non-zero) value of $\chi$ because they have been obtained on the basis of counting momentum dimensions. These relations suggest that $dL_3/dr=(4\chi+2d-2) \, C_3 \, \Lambda_0^{-4\chi-2d+2} \, e^{(4\chi+2d-2)r}$, $dL_4/dr=(5\chi+3d-2)\, C_4 \, \Lambda_0^{-5\chi-3d+2} \, e^{(5\chi+3d-2)r}$ and 
$dL_2/dr=(3\chi+d-2)\, C_2 \, \Lambda_0^{-3\chi-d+2} \, e^{(3\chi+d-2)r}$. These flow equations for $L_3(r)$, $L_4(r)$ and $L_2(r)$ contain unknown constants $C_3$, $C_4$ and $C_2$ respectively. The use of perturbation theory in our calculations serves to find these flow equations, along with the unknown constants, directly from the KPZ equation. In addition, we see that a good numerical input for $\chi$ results in good estimates for skewness and kurtosis values.

All recent numerical simulations in $(2+1)$ dimensions suggest 
that the roughness exponent $\chi$ is very close to $0.39$ which is close to $9/23$ ($0.391304$). 
We therefore slightly vary  the value of the roughness exponent to  $\chi =19/50$ ($=0.38$)
and $\chi= 2/5$ ($= 0.40$) and recalculate the integrals. We find that skewness 
and kurtosis values undergo shifts by less than $5\%$ from the calculated values 
given in Eqs. $(67)$ and $(69)$.

The estimated skewness and kurtosis values  of Chin and den Nijs \cite{Chin-Marcel99PRE.59.2633} 
via the Kim-Kosterlitz (KK) and BCSOS models are given in Table 1. Although their roughness exponents differed in the two 
models (KK: $\chi=2/5$ and BCSOS: $\chi\sim 0.38$), their skewness value ($|S|=0.27\pm0.01$) was the same for both models. 
Consequently, they concluded that the third moment is more reliable than the roughness exponent for a better identification 
of the universality class. Their kurtosis value was $Q=0.15\pm0.02$ for both models.

Reis \cite{PhysRevE.69.021610} considered the stationary states for etching, ballistic deposition, and body-centered restricted 
solid-on-solid (BCRSOS) models that suggested the universality of the absolute values of skewness and kurtosis. The best estimates 
come from etching model which yielded $|S|=0.26\pm 0.01$ and $Q=0.134\pm0.015$.

Miranda and Reis \cite{Miranda-Reis08_PRE.77.031134} used Euler discretization method for numerical integration of the KPZ equation
and obtained roughness exponent $0.37\leq \chi \leq 0.40$. In addition, they estimated skewness $S=0.25\pm 0.01$ and kurtosis $Q=0.15\pm0.1$ 
by extrapolating data in the limit of large substrate size $L$. Marinari et al.\ \cite{Marinari00_JPA} obtained skewness $S\sim 0.266$ and 
kurtosis $Q\sim 0.121$ through a numerical RSOS model.

Halpin-Healy \cite{Healy13_PRE.88.042118} considered the $(2+1)$-dimensional numerical models such as $g5_1$ DPRM, RSOS and KPZ Euler 
in the asymptotic limit of time ($t>>L^z$) and obtained a $(2+1)$-dimensional analog of 
$(1+1)$-dimensional Baik-Rains distribution from these numerical models. In addition, they calculated the Baik-Rains constant from those 
numerical models. On the other hand, instead of full probability distribution 
function only the skewness and kurtosis values have been estimated via the Kim-Kosterlitz (KK), BCSOS models \cite{Chin-Marcel99PRE.59.2633}, 
etching model \cite{PhysRevE.69.021610}, and KPZ Euler discretization approach \cite{Miranda-Reis08_PRE.77.031134}, as displayed in Table-I.

Experiments on vapor deposited oligmer thin film growth \cite{Palasantzas2002357,Tsamouras01-APL} yield the roughness and growth exponents 
$\chi=0.45\pm 0.04$ and $\beta=0.28\pm 0.05$, and the measured value of skewness $S=0.23$, suggesting that this growth is in the KPZ universality 
class. Halpin-Healy and Palasantzas \cite{Halpin-Healy-EPL-105-50001-2014} examined two point statistics, in particular,
spatial covariance by using the experimental results \cite{Palasantzas2002357}. In addition, they studied the local squared roughness distribution 
and extremal height distribution via Euler integration of $(2+1)$ KPZ and compared with the experimental results.

 Derrida and Appert \cite{Derrida99} (DA) defined a ratio 
$R_{DA}=S^2/Q$ in $1+1$ dimensions, called the Derrida-Appert ratio \cite{Healy-Takeuchi15}, and estimated the ratio as 
$R_{DA}=0.41517$, which is very close to the estimations from asymmetric simple exclusion principle (ASEP), BD and Brick models
in asymptotic time limit, suggesting that $R_{DA}$ is universal for the KPZ dynamics. In these dimensions, $R_{DA}$ is 
conjectured to be universal via the Derrida-Lebowitz universal scaling function (DLSF) which is independent of any model
parameters \cite{Chia-PRE-72-051102-2005}. Subsequently, Pr\"ahofer and Spohn \cite{Michael_Herbert_00} 
estimated skewness $S=0.35941$ and kurtosis $Q=0.28916$ for $(1+1)$-dimensional KPZ height fluctuations in the 
stationary state and thereby, DA ratio is calculated as $R_{DA}=0.44673$. Halpin-Healy and Takeuchi \cite{Healy-Takeuchi15} 
studied the higher dimensional numerical models of KPZ class and different geometrical sub-classes namely point-point, 
point-line and point-plane in the transient regime and found a approximate constant value of $R_{DA}$. 
Alves et al. \cite{Alves-PRE-90-020103-2014} studied the transient state RSOS model starting from the 
flat initial condition in higher dimensions $d=3,4,5,6$ and found $S \sim d^{0.46}$ and $Q \sim d^{0.92}$. This appears 
to suggest that $S^2/Q$ is independent of $d$, supporting the greater universality of Derrida-Appert ratio proposed in
\cite{Healy-Takeuchi15}, via their extensive examination of KPZ systems in the transient regime, across dimensions, 
as well as geometry. Our calculated skewness and kurtosis values yield $R_{DA}=0.41547$, the normalized values of 
which is compared with the 
other stationary value in Table 1.

\begin{table}[ht]
\caption{\label{Table-I} Stationary state values of Skewness and Kurtosis in $(2+1)$ dimensions.}
\begin{center}
\begin{tabular}{l|c|c|c|c|c}
\hline\hline
{\it System of study}   & {\it Method} & $|S|$ & $Q$ & $R_{DA}$ & {\it Reference}\\ 
\hline
$g5_1$ DPRM &   Numerical  & $0.240$    & $0.18$   & $0.32$ &\cite{Healy13_PRE.88.042118} \\
\hline
$2+1$ RSOS &  Numerical    & $0.256$    & $0.18$   & $0.364$ &\cite{Healy13_PRE.88.042118} \\
\hline
KPZ Euler &  Numerical     & $0.236$    &  $0.17$  & $0.328$ &\cite{Healy13_PRE.88.042118}\\
\hline
KPZ Equation & Numerical     & $0.25\pm 0.01$  &  $0.15\pm 0.1$ & $0.42\pm 0.31$  &\cite{Miranda-Reis08_PRE.77.031134} \\
\hline
Etching model& Numerical   & $0.26\pm 0.01$    & $0.134\pm0.015$ & $0.50\pm0.09$ &\cite{PhysRevE.69.021610}\\
\hline
KK and BCSOS   &  Numerical   & $0.27 \pm 0.01$   & $0.15 \pm 0.02$ & $0.49\pm0.08$  &\cite{Chin-Marcel99PRE.59.2633} \\
\hline
(2+1) RSOS (FIT-I) & Numerical & $0.2669 \pm 0.0004$ & $0.146 \pm 0.002$ & $0.488 \pm 0.008$ & \cite{Pagnani-Parisi-PRE-2015} \\    
\hline 
(2+1) RSOS (FIT-II) & Numerical &  $0.2657 \pm 0.0004$ & $0.145 \pm 0.001$ & $0.487 \pm 0.005$ &  \cite{Pagnani-Parisi-PRE-2015}                \\
\hline
Present work  &  Analytical    & $0.2879$     & $0.1995$ & $0.4155$ &   Eqs.\ \ref{skewness2D}, \ref{kurtosis2D}\\
\hline
\end{tabular}
\end{center}
\label{table:nonlin} 
\end{table}

The universality class of a dynamical system is an important statistical property. In the earlier studies on surface growth, the universality class 
used to be obtained from only the scaling exponents. 
In the last two decades, it has been realized that despite the same scaling exponents, the distribution functions can be entirely different 
due to different initial conditions corresponding to different sub-universality classes. Thus the distribution function contains 
more statistical information about the system than the scaling exponents.
The analytical calculation of the distribution function is hardly possible. 
To economize on the amount of calculations, one can calculate a few higher order cumulants of the distribution function, 
the normalized values of which can be used as identifiers of the universality classes.

\section*{Appendix}

The term $F(\vk,\vp)$ appearing in Eq. \ref{W3intefkp} is defined as
$$ F=(F_1 + F_2+F_3+F_4)/R$$ 

We define $a=37/23$, $b=41/23$, and $P=|\vk_1+\vk_2|$, so that
\begin{eqnarray*}
&& F_1(\vk_1,\vk_2)= 33 \ k_1^{4a} + 4 \ k_1^{3a} (47 \ k_2^{a} + 24 \ P^{a}) \nonumber \\
&& F_2(\vk_1,\vk_2)=3 (k_2^{a} + P^{a})^2 (11 k_2^{2a} +10 k_2^{a} P^{a} + 3 P^{2a}) \nonumber \\
&& F_3(\vk_1,\vk_2)= 2 k_1^{2a} (203 k_2^{2a} + 176 k_2^{a} P^{a}+51 P^{2a}) \nonumber \\
&& F_4(\vk_1,\vk_2)= 4 k_1^{a} (47 k_2^{3a} + 88 k_2^{2a} P^{a} + 53 k_2^{a} P^{2a} +12 P^{3a}) 
 \end{eqnarray*}

\begin{eqnarray*}
 R(\vk_1,\vk_2)= 256 k_1^{b} k_2^{b} (k_1^{a} +k_2^{a} + P^{a})^5
 \end{eqnarray*}

The function $T(\vk,\vp,\vq)$ appearing in Eq. \ref{h4-Tkpq} is expressed as 
\begin{eqnarray*}
 T=(T_1+ T_2+T_3+T_4+T_5+T_6)/S
\end{eqnarray*}
where
\begin{eqnarray*}
 && T_1(\vk,\vp,\vq)= 3 \{ 33 k^{6a}+ 2 k^{5a} (127 p^{a} + 127 q^{a} +81 Q^{a})\}
 \end{eqnarray*}

\begin{eqnarray*}
T_2(\vk,\vp,\vq)=3[ k^{4a} \{815 p^{2a} + 815 q^{2a} +986 q^{a} Q^{a}  +327 Q^{2a} + 34 p^{a} (51 q^{a} + 29 Q^{a})\}]
 \end{eqnarray*}
\begin{eqnarray*}
&& T_3(\vk,\vp,\vq) = 3[4 k^{3a} \{297 p^{3a} + 297 q^{3a} +585 q^{2a} Q^{a} + 375 q^{a} Q^{2a}+ 87 Q^{3a} \nonumber \\
 && + 3 p^{2a} (397 q^{a}+ 195 Q^{a})+p^{a} (1191 q^{2a} + 1234 q^{a} Q^{a} + 375 Q^{2a})\}]
\end{eqnarray*}
 
\begin{eqnarray*}
&&T_4(\vk,\vp,\vq)= 3 [ (p^{a} + q^{a}+ Q^{a})^2 \{(33 p^{4a}+4 p^{3a} (47 q^{a} + 24 Q^{a})\nonumber  \\
&&  + 3 (q^{a} + Q^{a})^2 (11 q^{2a} + 10 q^{a} Q^{a} + 3 Q^{2a}) + 2 p^{2a} (203 q^{2a} + 176 q^{a} Q^{a}+ 51 Q^{2a})\nonumber \\
&&  + 4 p^{a} (47 q^{3a} + 88 q^{2a} Q^{a} + 53 q^{a} Q^{2a} + 12 Q^{3a})\}] \nonumber
\end{eqnarray*}

\begin{eqnarray*}
&&T_5(\vk,\vp,\vq)= 3[k^{2a} \{815 p^{4a} + 12 p^{3a} (397 q^{a} +195 Q^{a})+ ( q^{a} + Q^{a})^2  \nonumber \\
&& (815 q^{2a} + 710 q^{a} Q^{a} + 207 Q^{2a})+2 p^{2a} (5277 q^{2a}+4366 q^{a} Q^{a} + 1221 Q^{2a}) \nonumber \\
&& + 4 p^{a} (1191 q^{3a} + 2183 q^{2a} Q^{a} +1273 q^{a} Q^{2a} + 281 Q^{3a})\}]
\end{eqnarray*}

\begin{eqnarray*}
&& T_6(\vk,\vp,\vq)= 3 [2 k^{a} \{127 p^{5a} + 17 p^{4a} (51 q^{a} +29 Q^{a})+ (q^{a} + Q^{a})^3 (127 q^{2a} + 112 q^{a} Q^{a} + 33 Q^{2a})\nonumber \\
&& + p^{a} (q^{a} + Q^{a})^2 (867 q^{2a} + 734 q^{a} Q^{a}+ 211 Q^{2a}) + p^{3a} (2382 q^{2a} + 2468 q^{a} Q^{a}+ 750 Q^{2a}) \nonumber \\
&& + p^{2a} (2382 q^{3a}+4366 q^{2a} Q^{a} + 2546 q^{a} Q^{2a} + 562 Q^{3a})\}] \nonumber \\
\end{eqnarray*}
and 
\begin{eqnarray*}
 S(\vk,\vp,\vq)=4096 k^{4a} p^{4a} q^{4a} ( k^{a} + p^{a} + q^{a} + Q^{a})^7
\end{eqnarray*}
where $Q=|\vk+\vp+\vq|$

\section*{Acknowledgments}

T.S. is thankful to the Ministry of Human Resource Development (MHRD), Government of India, for financial support 
through a scholarship. M.K.N. is indebted to the Indian Institute of Technology Delhi, and particularly to 
Prof. Ravisankar and Prof. Senthilkumaran, for hospitality at I.I.T.  Delhi.

\end{document}